\documentclass[preprint,12pt,authoryear]{elsarticle}
\usepackage{amssymb}
\let\cite\citep
\usepackage{amsmath}
\usepackage[utf8]{inputenc}
\usepackage{graphicx}
\usepackage[labelfont=bf]{caption} 
\usepackage{subcaption} 
\usepackage{xcolor}
\usepackage{booktabs} 
\usepackage{tabularx}
\usepackage{ulem}
\usepackage{comment}
\usepackage{soul}
\usepackage{float} 
\usepackage{lineno}

\journal{Advances in Space Research}

\begin{document}

\begin{frontmatter}

\title{Conceptual Design of a Compact Magnetic Spectrometer for Space Radiation Measurement}
\author[SDU]{Y. Zhu} 
\author[SDIAT]{F. Faldi} 
\author[SDIAT]{V. Mikhailov} 
\author[SDIAT]{Z. Qu} 
\author[SDU,SDIAT]{A. Stuzhin} 
\author[SDU,SDIAT]{W. Xu} 

\cortext[cor1]{Corresponding author:
zhaoyi.qu@iat.cn; weiwei.xu@iat.cn}  

\affiliation[SDU]{organization={Shandong University (SDU)},
            addressline={250100}, 
            city={Jinan},
            state={Shandong},
            country={China;},
            mail={\newline~yiheng.zhu@mail.sdu.edu.cn(Y.Z),aleksandr.stuzhin@cern.ch(A.S)}}
\affiliation[SDIAT]{organization={Shandong Institute of Advanced Technology (SDIAT)},
            addressline={250100},
            city={Jinan},
            state={Shandong},
            country={China;},
            mail={\newline~zhaoyi.qu@iat.cn(Z.Q),weiwei.xu@iat.cn(W.X),vladimir.mikhailov@iat.cn(V.M),\newline francesco.faldi@iat.cn(F.F)}}

\begin{abstract}
We present the design of COMIRA, a compact magnetic spectrometer for space radiation measurement. The spectrometer consists of four layers of scintillating fiber trackers, four layers of silicon strip trackers, one threshold Cherenkov counter, two anti-coincidence counters, and a permanent magnet.
The detector has an overall dimension of approximately $12 \times 12 \times 12~\mathrm{cm}^3$ and a geometrical acceptance of \(\sim 6.6~\mathrm{cm}^2\,\mathrm{sr}\).
Based on combined measurements of charge, velocity, and rigidity (momentum/charge), it enables the simultaneous measurement of the energy spectrum, angular distribution and time evolution of individual radiation components, including electrons, positrons, protons and light nuclei over an energy range from tens of MeV to several GeV.
The design can be further optimized depending on the available resources and constraints of the hosting space platform.
\end{abstract}




\end{frontmatter}


\section{Introduction}
\label{Introduction}
Characterizing the particle radiation environment in Low Earth Orbit (LEO) is essential not only for understanding the mechanisms governing the acceleration and transport of charged particles in near-Earth space, but also for assessing radiation hazards affecting spacecraft operations and human safety during space missions. 

The LEO environment is continuously exposed to a complex mixture of charged particles, which includes primaries from Galactic Cosmic Rays (GCR) and Solar Energetic Particle (SEP), and secondaries generated by the interaction of primaries with residual atmosphere. Trapped particles within the Earth's radiation belts and SEP events represent the dominant contributions to the radiation exposure experienced by spacecraft and astronauts in LEO.

Existing compact radiation monitors, designed primarily for solar physics applications, often lack the energy resolution and particle identification capability required to disentangle the complex mixture of particle species and energies present in the low-altitude radiation environment. The main advantage of a magnetic spectrometer, in this context, is the ability to precisely measure the energy spectra of dominant charged particle populations entering the magnetosphere, while also providing high sensitivity to less abundant components, including antimatter particles, through charge-sign discrimination.

\begin{figure}[htb]
   \centering
      \hspace*{-0.5 cm}
\includegraphics[width=0.9\textwidth]{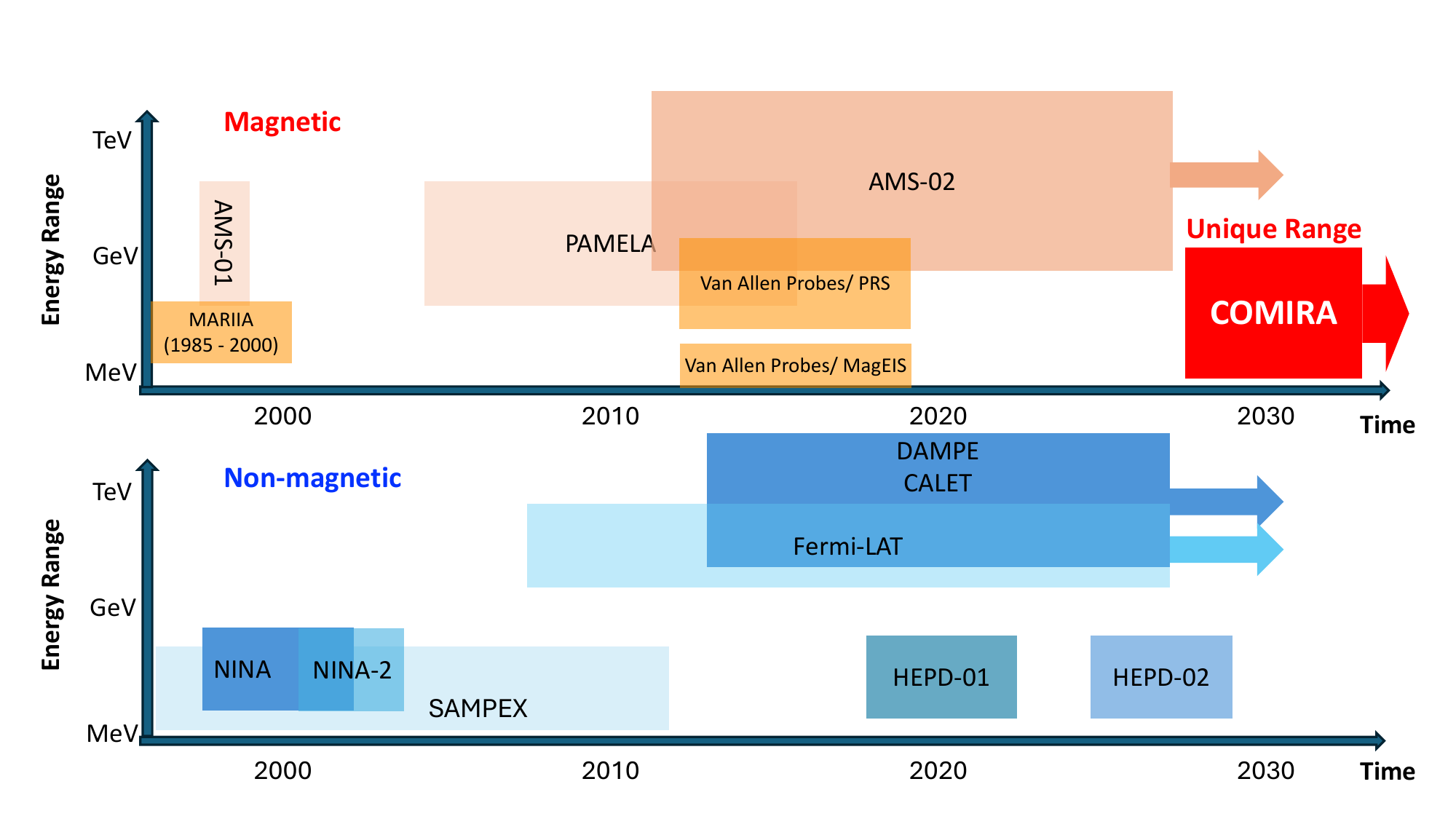}
    \caption{\footnotesize The timeline and energy coverage of recent experiments measuring high-energy fluxes of charged particles in LEO. This work proposes a magnetic spectrometer, COMIRA, to investigate a unique energy range from tens of MeV to several GeV.}
  \label{fig:Review_Experiments}
\end{figure}

Figure~\ref{fig:Review_Experiments} summarizes the recent magnetic and non-magnetic experiments which measure charged particles in near-Earth space.
High-precision magnetic spectrometers, such as AMS-01 \cite{ams01}, PAMELA \cite{pamela}, and AMS-02 \cite{ams02}, have provided measurements of cosmic ray fluxes and composition above GeV energies.
Magnetic spectrometry had been applied to magnetospheric radiation measurements in the MeV energy range.
MARIIA on the Salyut-7 station in 1985 and MARIIA-2 on the Mir station from 1987 to 2000 measured electrons and positrons between 20 and 150 MeV \cite{MARIIA_2003}.
The Van Allen Probes mission was operational between 2012 and 2019 \cite{Spence2013-ze}: its Magnetic Electron Ion Spectrometer (MagEIS) measured trapped electrons in the energy range from 20 keV to 4.8 MeV and protons from 55 keV to 20 MeV \cite{Blake2013-vp}, while the Radiation Belt Storm Probes Ion Composition Experiment (RPS) measured protons between 50 MeV and 2 GeV \cite{Mazur2023-zi}. 

There are also many non-magnetic experiments for measuring cosmic rays in space without distinguishing the charge sign. At high energies, DAMPE \cite{Chang2024-kq} and CALET \cite{Akaike2024-cl} can measure fluxes of protons, nuclei, and (electrons+positrons) beyond the TeV range. 
Examples of calorimeters that measured cosmic ray fluxes up to hundreds of MeV include: SAMPEX \cite{Baker_1993_SAMPEX}, which measured electrons and ions composition in a near-polar orbit from 1992 to 2012. NINA and NINA-2 \cite{Bakaldin1997-vk} measured light nuclei, isotopes, and the solar component up to 200 MeV/n in the early 2000s. HEPD-01 \cite{hepd} and HEPD-02 \cite{Bartocci2026-bz} on board the CSES satellites measure the charged component of cosmic and trapped radiation up to 200 MeV.

A compact magnetic spectrometer, suitable for deployment on small satellites, could provide unique measurements of magnetospheric radiation components, covering the energy gap in the observation between tens of MeV, typically explored by solar and space physics missions, and hundreds of MeV measured by high energy cosmic ray instruments.
A recently proposed concept is the PAN experiment \cite{Wu2019-zz}, a version of which (pix.PAN) is designed for the COMPASS mission to study charged particle fluxes in the high-rate environment of Jupiter's radiation belts \cite{Hulsman2023-gk}. Its energy measurements span between 10 MeV and 1 GeV for electrons and between 200 MeV and a few GeV for protons. A demonstrator (mini.PAN) has already been built and tested \cite{Silvestre2024-ac}.


In this work, we present the reference design of COMIRA, a compact magnetic spectrometer, which determines the charge, velocity, and rigidity of charged particles. By operating in LEO, COMIRA is able to measure the energy spectrum and angular distribution of individual radiation components over an energy range from tens of MeV to several GeV. The simultaneous measurements of electrons, positrons, protons, and light nuclei would provide unique information on the origin and dynamics of the radiation in near-Earth space.

\section{Scientific Objectives}
\label{Scientific Objectives}


A major fraction of charged radiation in LEO is concentrated within the South Atlantic Anomaly (SAA), which is part of the inner radiation belt. Charged particles are trapped and drift in the geomagnetic field, bouncing between magnetic polar regions \cite{walt2005introduction}. The inner belt is mainly formed by protons and electrons up to several hundred MeV and electrons up to MeV.
Protons are mainly generated via the CRAND mechanism by neutrons decayed inside the belt \cite{selesnick2023inner} while electrons are produced with lower energy \cite{li2022inner} and can also be directly produced in the interaction of GCRs with the residual atmosphere.
The viability and efficiency of the CRAND mechanism as an electron source can be confinement by measurements of the antimatter component in the MeV to GeV energy range, which can be separated effectively by COMIRA's magnet. As discussed in \cite{Galper1996} a large population of trapped electrons inside the inner Van Allen Belt is expected, while the positron counterpart is thought to be in agreement with secondary origin, produced in atmospheric interactions.
One of the primary scientific goals of this instrument is to perform a continuous high-resolution spatial and spectral mapping of the SAA, providing a complementary spectrometric measurement to similar calorimetric counterparts \cite{Bartocci2025-et}, including the antimatter component, which can be separated effectively by the magnet. 

Trapped radiation is also characterized by anisotropic distribution, which can be explored only with a tracking detector \cite{bruno2021east}.
By studying the pitch angle distribution, COMIRA could put constraints on the acceleration mechanism and particle losses inside the magnetosphere in this energy range.

COMIRA could be suited to study the presence of partially ionized atoms in GCRs, also known as \textit{Anomalous Cosmic Rays} (ACR) \cite{Giacalone2022-pn}.
The observation of ACRs at the energies measured by COMIRA could constrain the maximum energy at which these ions are accelerated, providing an input on the determination of the mechanism responsible for this acceleration, while also providing an additional measurement of their composition.

The LEO radiation environment is highly sensitive to transient solar energetic events, which generate geomagnetic storms and drive SEPs up to several GeV \cite{Reames2021-sf}.
The study of the ionic charge state of SEPs, achievable with COMIRA through the geomagnetic cutoff technique \cite{Mazur1999-oh}, and energy spectra could improve the understanding of their origin and acceleration mechanisms.

Additionally, the arrival of SEP protons and ions has been shown to be anticipated by relativistic electrons \cite{Posner2007-sz}, which act as a precursor for the event, providing an interesting use case for COMIRA to forecast SEP events.

\section{Reference Conceptual Design}

\begin{figure}[!htb]
    \centering
    \begin{subfigure}[b]{0.45\textwidth}  
        \centering  
        \includegraphics[width=\linewidth]{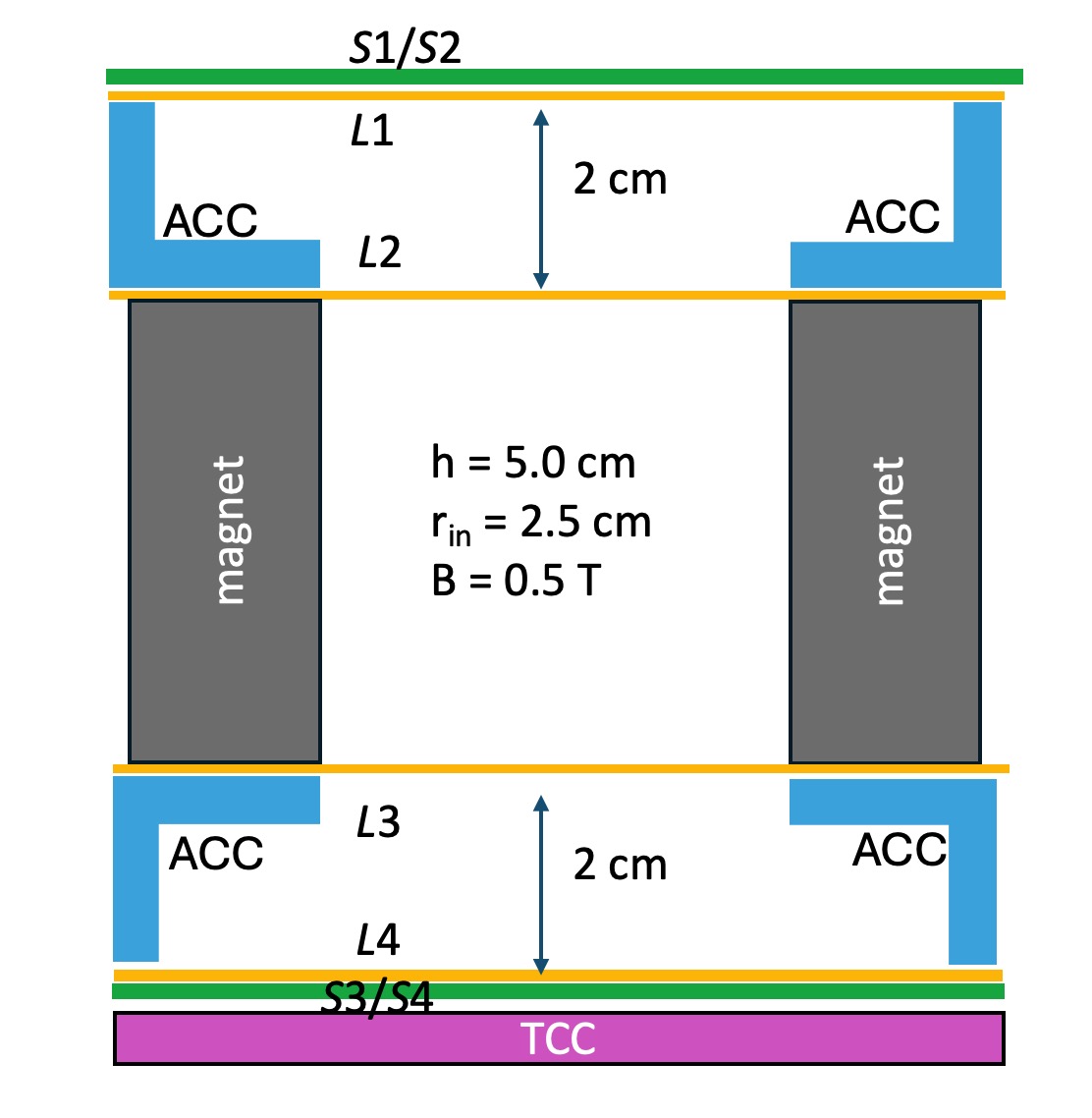}     
        \label{fig:left}
    \end{subfigure}
    \hfill
    \begin{subfigure}[b]{0.45\textwidth}
        \centering
        \includegraphics[width=\linewidth, trim=8.4cm 2.4cm 2.9cm 1.1cm, clip]{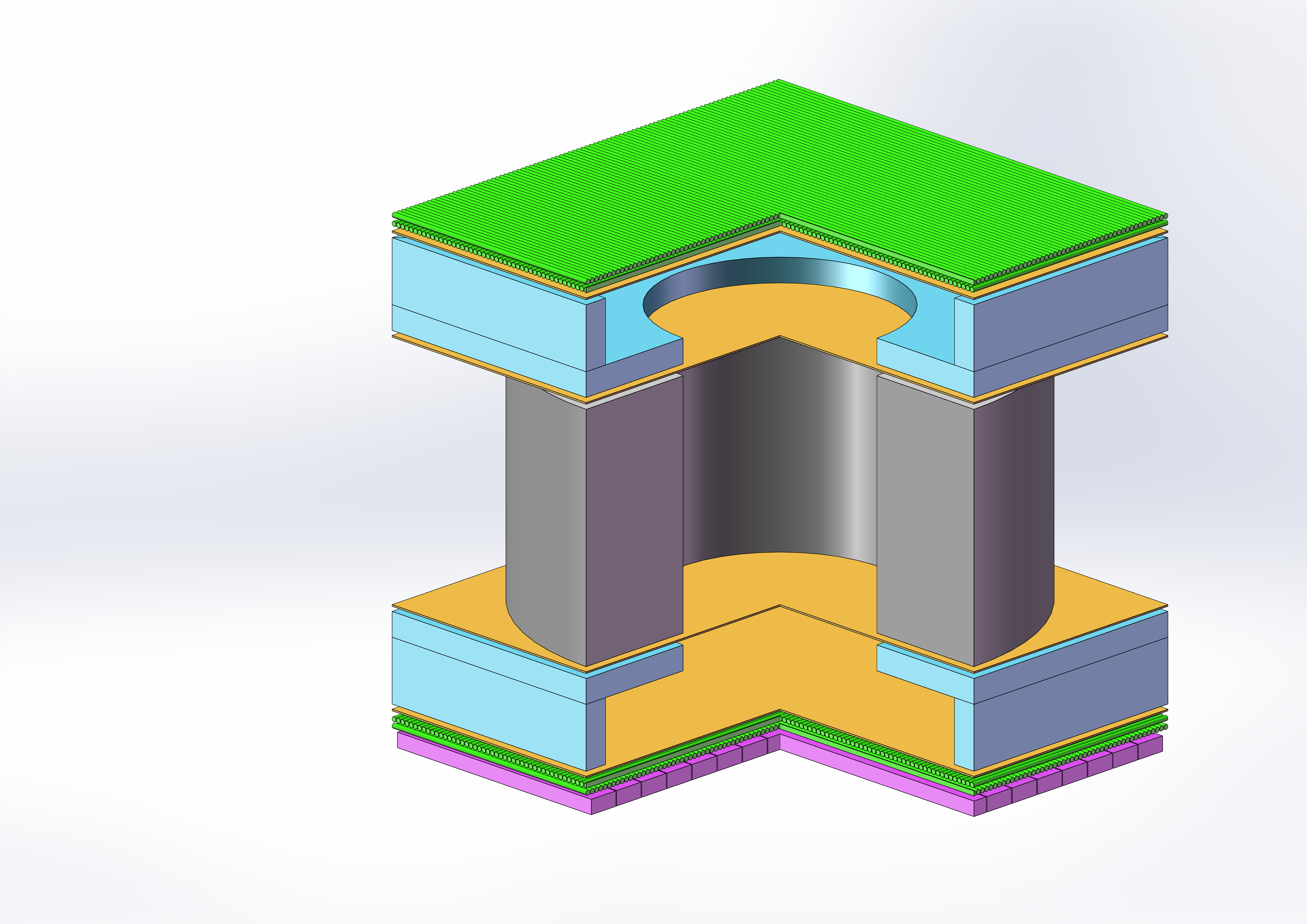}    
        \label{fig:right}
    \end{subfigure}
    
    \caption{The schematic layout (left) and three-dimensional view (right) of the compact magnetic spectrometer COMIRA. The spectrometer consists of the scintillating fiber tracker (SciFi, green, S1--S4), silicon strip tracker (STK, yellow, L1--L4), threshold Cherenkov counter (TCC, violet), anti-coincidence counter (ACC, blue), and permanent magnet (grey).}
    \label{fig:CubeSAT-layout}
    
\end{figure}

The baseline detector design is developed under the assumption of a small satellite payload with an approximate envelope of \(12\times12\times12~\mathrm{cm}^3\). 
The spectrometer consists of four layers of scintillating fiber tracker (SciFi), four layers of double-sided silicon strip tracker (STK), two barrel-shaped anti-coincidence counters (ACC) made of plastic scintillators, one threshold Cherenkov counter (TCC), and one permanent magnet. 
The schematic layout of the detector is shown in Figure ~\ref{fig:CubeSAT-layout}.
The detector specifications and required performance are summarized in Table~\ref{table:detector}.

\begin{table}[!htbp]
\centering
\caption{Specifications and performance requirements of the sub-detectors.}
\small
\renewcommand{\arraystretch}{1.15}
\begin{tabular}{p{3.5cm}|p{5.5cm}|p{3cm}}
\hline
\textbf{Detector} & \textbf{Specifications} & \textbf{Requirements} \\
\hline
Scintillating Fiber Tracker &
10 cm fiber mat; 4 staggered layers of 250 $\mu$m diameter fibers; 6 mm pitch SiPM at both ends.
&
time resolution $\Delta^{t}_{\textrm{SciFi}}=80$ ps
\\
\hline
Silicon Strip Tracker &
$10\times10$ cm$^2$ area; 320 $\mu$m thick; 65 $\mu$m pitch; 1:3 strip readout; double sided sensor
&
position resolution $\sigma_{\textrm{SSD}}=15~\mu$m
\\
\hline
Threshold Cherenkov Counter &
NaF bar $100\times3\times6.2$ mm$^3$, refactive index $n_{\textrm{NaF}}=1.33$; $6\times6$ mm$^2$ SiPM readout
&
Proton rejection below 1 GeV/c
\\
\hline
Anti-coincidence Counter &
10 cm long, "L"-shape cross section, plastic scintillator
&
$>99\%$ efficiency for $Z=1$
\\
\hline
Permanent Magnet &
2.5 cm inner radius, 5 cm height; Halbach configuration with high-grade NdFeB segments 
&
0.5 T dipole field
\\
\hline
\end{tabular}
\label{table:detector}
\end{table}

The magnet is located at the center of the payload to bend the trajectories of incident charged particles. 
In the baseline design, a cylindrical magnet with an inner radius of \(R_{\mathrm{in}}=2.5\)~cm and a height of 5~cm is adopted. 
Using the commonly employed Halbach configuration, the dipole magnet is constructed from an array of 16 high-grade NdFeB segments. 
For typical modern NdFeB permanent magnets, the remanent magnetic field is \(B_r=1.2\sim1.4\)~T. 
According to the relation \(B = B_r \cdot \ln(R_{\mathrm{out}}/R_{\mathrm{in}})\), a dipole field of \(B=0.5\)~T can be achieved with an outer radius of \(R_{\mathrm{out}}\approx4\)~cm.

The detector coordinate system is concentric with the center of the magnet. The $x$ axis is parallel to the main component of the magnetic field, and the $z$ axis points vertically to the local zenith. The $(y-z)$ plane is the bending plane. Terms such as ‘‘above,’’ ‘‘below,’’ ‘‘downward-going’’, and ‘‘upward-going’’ refer to the detector coordinate system.

Four layers of scintillating fiber (SciFi) detectors are placed at the outermost region of the spectrometer, with two layers on the top and two layers at the bottom. 
Each SciFi module consists of four staggered layers of 250~\(\mu\)m-diameter scintillating fibers coupled at both ends to arrays of silicon photomultipliers (SiPMs), whose pixel pitch is 6~mm.
Each module measures the hit coordinate of charged particles in the direction perpendicular to the fibers with a spatial resolution of 1.7~mm 
and provides a timing resolution of 80~ps for traversing particles.
The first layer, \(S1\), is located at \(z=5\)~cm with fibers aligned along the \(x\)-axis, measuring the \(y\)-coordinate. 
The second layer, \(S2\), is located at \(z=4.5\)~cm with fibers orthogonal to \(S1\), measuring the \(x\)-coordinate. 
The two bottom layers, \(S3\) and \(S4\), are placed at \(z=-4.5\)~cm and \(z=-5\)~cm, measuring the \(y\)- and \(x\)-coordinates, respectively.
With timing information from the four SciFi layers, it determines the upward- and downward-going direction of charged particles and provides a measurement of their velocity.

Four layers of double-sided silicon strip detectors (SSDs) form the precision tracker (STK). 
The first layer, \(L1\), is located below the SciFi layer \(S2\) at \(z=4.5\)~cm. 
Two SSD layers are placed symmetrically above and below the magnet, with \(L2\) at \(z=2.5\)~cm and \(L3\) at \(z=-2.5\)~cm. 
The fourth layer, \(L4\), is located above the SciFi layer \(S3\) at \(z=-4.5\)~cm.
Each SSD layer provides an active area of \(10\times10~\mathrm{cm}^2\) and measures the coordinate of charged particles in both the bending and non-bending planes. 
The SSD employs a 65 µm strip pitch with a 1:3 readout scheme (one readout strip followed by three floating strips). Together with analog charge centroid reconstruction, a spatial resolution of $10 - 15$~\(\mu\)m is achieved while reducing the number of readout channels by a factor of four.
The particle trajectory in the bending direction is reconstructed using combined measurements from the SciFi and STK systems, yielding a lever arm of up to 10 cm.

The Threshold Cherenkov Counter (TCC) is located at the bottom of the spectrometer. 
The TCC consists of an array of modules, each comprising a NaF crystal bar with dimensions of 10~cm in length, 0.3~cm in thickness, and 0.62~cm in width. 
The NaF crystal surfaces are coated with tetraphenyl butadiene (TPB) wavelength shifter and wrapped with an enhanced specular reflector (ESR) film, providing a reflection efficiency of 98.5\%. 
A silicon photomultiplier (SiPM) is coupled to one end of each NaF bar to collect Cherenkov photons. 
The refractive index of NaF is \(n_{\mathrm{NaF}} \simeq 1.33\), so charged particles with velocity \(\beta > 1/n_{\mathrm{NaF}}\) produce Cherenkov radiation. 
The TCC therefore rejects particles with \(\beta < 1/n_{\mathrm{NaF}}\), corresponding to a proton momentum threshold of approximately 1~GeV/\(c\).

The trigger is primarily formed by signals from the SciFi tracker and the anti-coincidence counter (ACC), selecting events that traverse all four SciFi layers (\(S1 \,\&\&\, S2 \,\&\&\, S3 \,\&\&\, S4\)) while producing no signal in the ACC (\(N_{\mathrm{ACC}} = 0\)). 
The threshold Cherenkov counter (TCC) provides an additional capability to suppress low-energy protons and nuclei (\(< 1~\mathrm{GeV}/c\)) by requiring the detection of Cherenkov photons. 
A dynamic trigger logic can be implemented by incorporating the TCC response depending on the instantaneous event rate.

The spectrometer is designed to be nearly symmetric, enabling the detection of particles entering from both the top and bottom. 
Time-of-flight information provided by the SciFi system is used to distinguish downward-going and upward-going particles. 
By combining the magnetic deflection measured by the SciFi and STK, the charge sign of each particle can be determined, enabling separation of matter and antimatter, such as electrons from positrons and antiprotons from protons.

The evaluation of weight, power consumption, data volume, telemetry bandwidth, and other specifications of the COMIRA will be presented in the forthcoming engineering design report.

\section{Detector Simulation and Event Reconstruction}

A Monte Carlo (MC) simulation program has been developed based on the GEANT4.10.5.1 \cite{Agostinelli2003,allison2006,allison2016recent} to perform a detailed simulation of detector responses.
This program includes the description of the compact magnetic spectrometer and simulates electromagnetic and hadronic interactions of particles in the materials of the spectrometer. 
The MC truth information, including the trajectory of particle transportation, the time at which particles hit the sensitive volume of the detectors, and the energy deposition in the sensitive volume, is recorded.
The detector responses are then generated by a digitization process that uses the MC truth information and accounts for the typical characteristics of the detector components.

For the SciFi, the measured coordinate and time at each layer is generated from a Gaussian sampling. The mean value of the sampling function is taken from the MC truth coordinate and time of the simulated particle at the specific layer, and sigma is $\sigma_\textrm{SciFi} = 1.7 ~\mathrm{mm}$ and $\Delta^{t}_{\textrm{SciFi}} =80 ~\textrm{ps}$, respectively.  

For the STK, the measured coordinate at each layer is generated from the same Gaussian sampling procedure with coordinate resolution of the SSD $\sigma_{\textrm{SSD}}=15 ~\mu\mathrm{m}$.

For the ACC, the energy depositions in upper and lower barrels are effectively converted to the total number of photoelectrons generated in the SiPM, $N^{\mathrm{ACC}1}_\mathrm{p.e.}$ and $N^{\mathrm{ACC}2}_\mathrm{p.e.}$. During the conversion, the SiPM photon detection efficiency and photon propagation efficiency inside the plastic scintillator have been accounted for. 

For the TCC, the generated Cherenkov photons are propagated in the NaF crystal and reflected at its surface. The total number of photoelectrons generated in the SiPM, $N^{\mathrm{TCC}}_\mathrm{p.e.}$, accounting for SiPM photon detection efficiencies, is recorded and used for particle identification and possible integration into the trigger logic.

To simulate the trigger, the energy deposition measured in four layers of SciFi and two barrels of ACC is used. In the following performance study, we required that all the four SciFi passing the threshold of $\Delta E_{\textrm{SciFi}}>0.25 \times \Delta E^{\textrm{MIP}}_{\textrm{SciFi}}$ and all the ACC counters below the threshold $\Delta E_{\textrm{ACC}}<0.25 \times \Delta E^{\textrm{ACC}}_{\textrm{MIP}}$.

The simulated events then undergo a reconstruction program. 
The trajectory is built from the coordinates in four SciFi layers ($S1, S2, S3$, and $S4$) and four STK layers ($L1, L2, L3$, and $L4$). 
The rigidity (momentum/charge, $R=p/Z$) of the charged particle is measured from the angular difference before and after the particle enters the magnet. 
The velocity $\beta$ is reconstructed from the time measurements of four SciFi layers and the path length of the trajectory. The upward ($\beta<0$) or downward ($\beta>0$) direction of the charged particle is determined by the time difference of the four SciFi layers. 
The charge sign of the particle is then determined from the sign of velocity and rigidity. 

\section{Expected Performance}
The performance of the designed compact magnetic spectrometer has been evaluated using the MC simulation. The results on the geometrical acceptance, rigidity resolution, velocity resolution, and particle identification will be presented. 

\subsection{Geometrical Acceptance}

The geometrical acceptance is evaluated using the MC simulation in which the particles are generated from a sphere surrounding the detector. Isotropic and uniformly distributed particles are generated from the inner surface of the sphere. 
The detector acceptance is calculated as the product of the sphere area and the ratio of triggered events to the total number of simulated events.

\begin{figure}[!htb]
   \centering
    \includegraphics[width=0.75\textwidth]{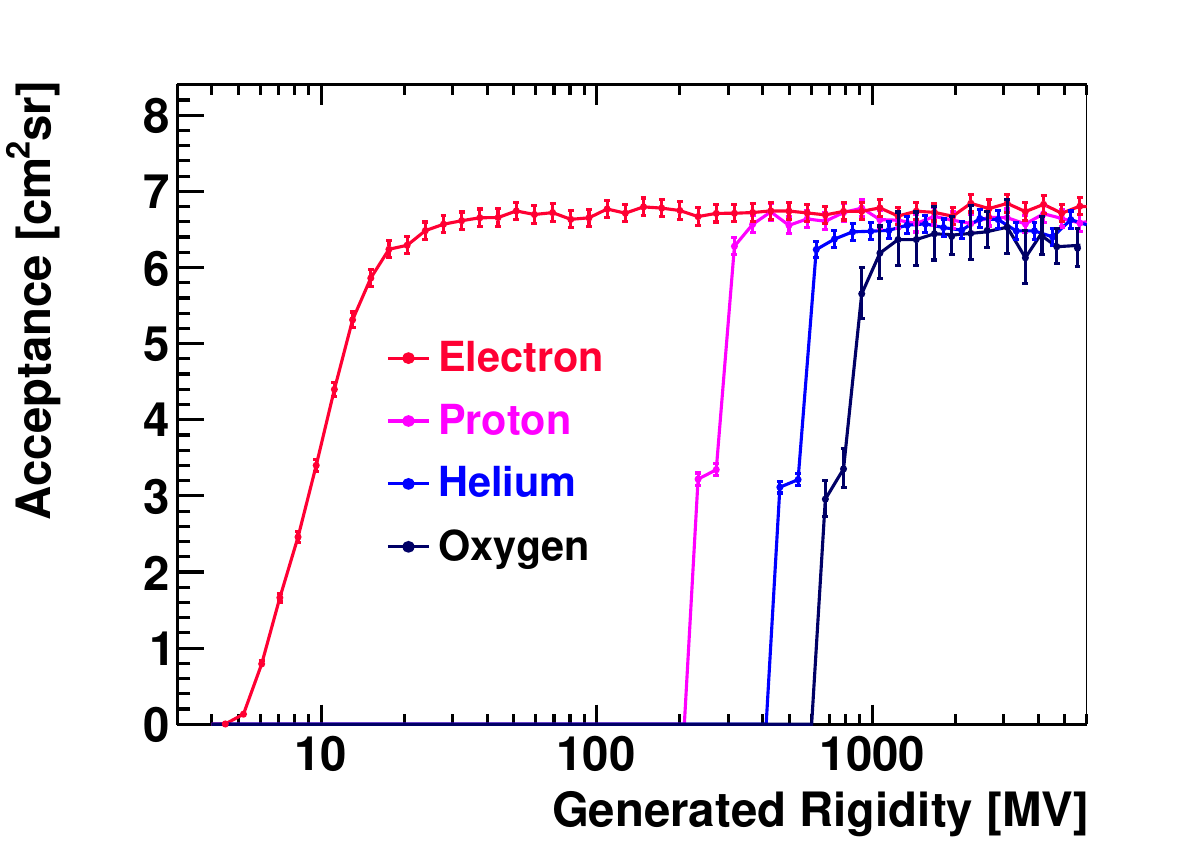}
    \caption{\footnotesize The geometrical acceptance of the compact magnetic spectrometer for electrons and positrons (red), protons (magenta), helium (blue) and oxygen (black).} 
  \label{fig:acceptance}
\end{figure}

The simulated events are required to pass through four layers of SciFi, four layers of STK, and the inner bore of the magnet, while not passing the threshold in ACC. Due to the symmetric geometry of the detector, both the downward- and upward-going particles are collected.  
Figure \ref{fig:acceptance} shows the dependence of the acceptance for electrons, protons, helium and oxygen on the generated rigidity. As seen, the acceptance is approaching  $6.6~\textrm{cm}^{2}\textrm{sr}$ for electrons and positrons above $\sim15$ MV, $\sim230$ MV for protons, $\sim460$ MV for helium nuclei, and $\sim 1$ GV for oxygen nuclei. The reduction of geometrical acceptance at lower rigidities is due to the energy loss in the detector material.

\subsection{Rigidity Resolution and Energy Range}
The three dimensional trajectory of the charged particles are measured by the SciFi and STK. 
As seen from Figure~\ref{fig:CubeSAT-layout},  
The SciFi layers $S1$ and $S4$ measure the bending ($y$) coordinates, $S2$ and $S3$ the non-bending ($x$) coordinates. 
The STK layers $L1-L4$ measure ($x-y$) coordinates at four $z$ positions. 
Together, the trajectory projections in $x-z$ and $y-z$ plane are each measured with six points, three before and three after entering the magnet. 

\begin{figure}[!htb]
   \centering
    \includegraphics[width=0.75\textwidth]{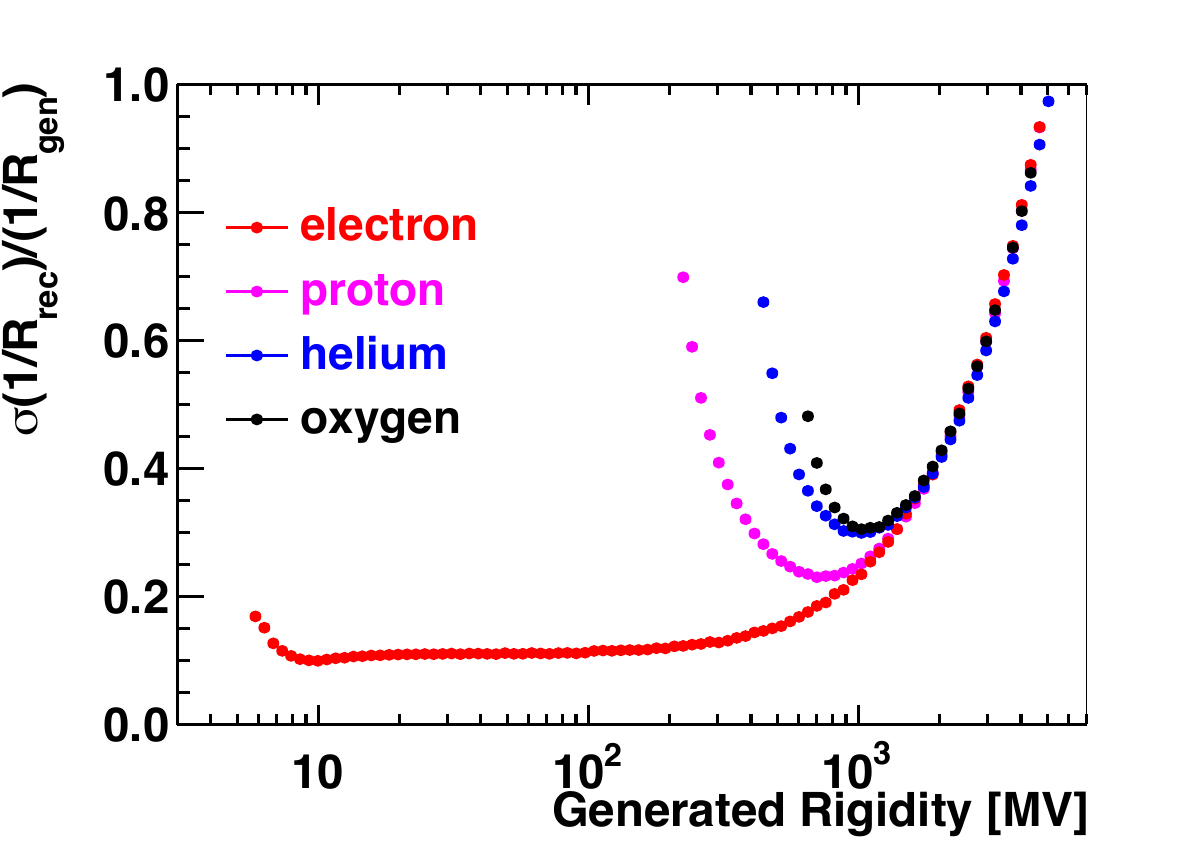}
    \caption{\footnotesize The rigidity resolution as a function of generated rigidity for electrons (red), protons (pink), helium (blue) and oxygen (black). As seen, the maximum detectable rigidity is $\sim5$ GV. }
  \label{fig:rigidity_resolution}
\end{figure}

The rigidity of the charged particle is reconstructed from the change of the trajectory slope in the bending plane caused by the magnetic field.  
As shown in Figure \ref{fig:rigidity_resolution}, the intrinsic rigidity resolution of the spectrometer is plotted, with an estimated MDR (maximum detective rigidity) of $\sim5$ GV. 

The measured rigidity can be then converted to the energy or kinetic energy. 
For electrons, the designed spectrometer covers an energy range from $\sim5$ MeV to 5 GeV; for protons, it covers the kinetic energy range from $\sim30$ MeV/n to 4 GeV/n; 
for $^4\textrm{He}$ and $^{16}\textrm{O}$, it covers the kinetic energy range from $\sim30$ MeV/n to 1.7 GeV/n.
For positrons, the energy coverage is determined by the proton rejection power and electron charge confusion level, and will be discussed in the following sections. 

\subsection{Velocity Resolution and Direction Determination}
The velocity ($\beta = v/c$) vector is reconstructed from the time measurements by four SciFi layers and the reconstructed trajectory using the time-of-flight (TOF) technique.  

With four measurements at a level arm of 10 cm and single layer time resolution of 80 ps, the probability to mis-reconstruct particle direction, i.e. identifying an downward-going particle as upward-going, is below $3.8\times10^{-5}$.

\begin{figure}[!htb]
   \centering
    \includegraphics[width=0.75\textwidth]{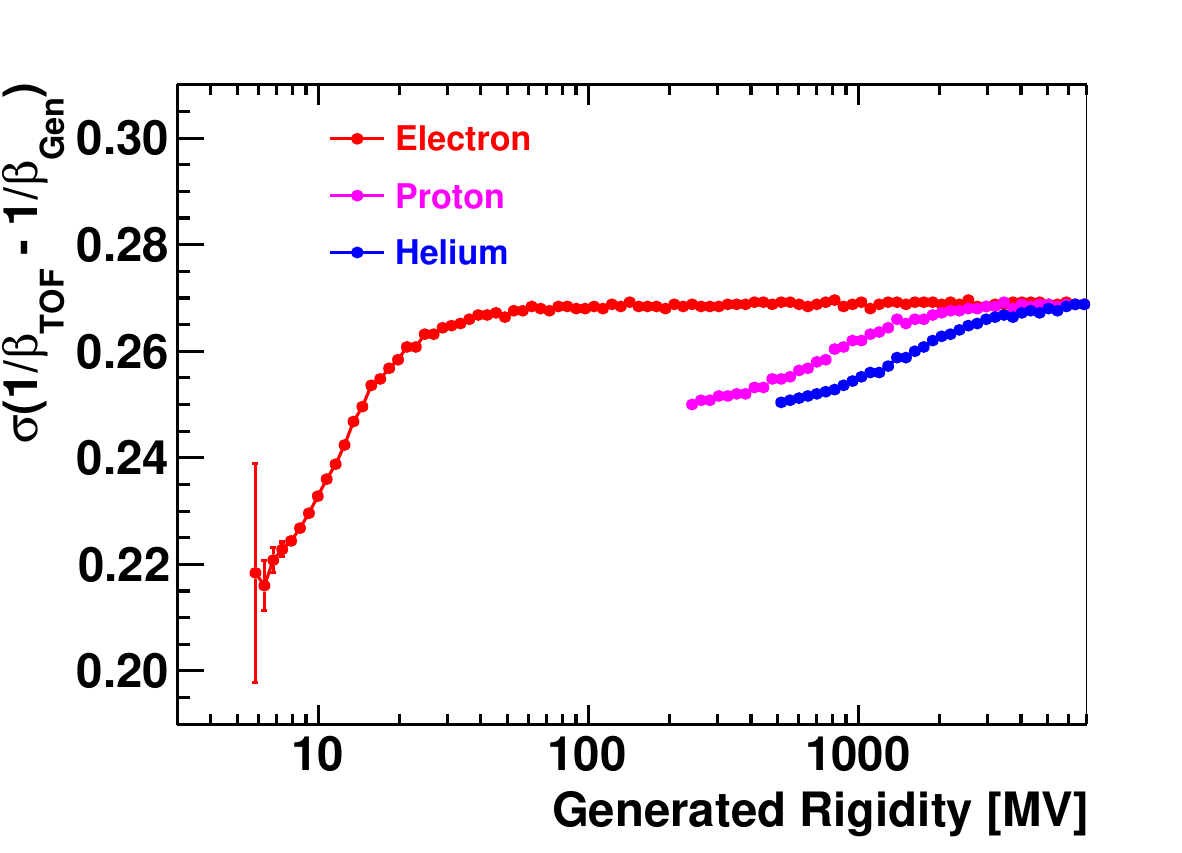}
    \caption{\footnotesize The velocity resolution of reconstructed with TOF as a function of generated rigidity for electrons (red), protons (pink), and helium (blue). 
    }
  \label{fig:beta_resolution}
\end{figure}

The resolution of $\beta_\textrm{TOF}$ from time-of-flight (TOF) technique is determined by the time resolution and the precision of path-length integration.
Figure \ref{fig:beta_resolution} shows the $\beta$ resolution. As seen, the $\beta_\textrm{TOF}$ resolution is better than $27\%$.

\subsection{Field of view and angular resolution}

The incident particle direction in the local detector coordinate system is determined through trajectory extrapolation to the top or bottom of the instrument. 
For a downward-going particle, the extrapolated trajectory direction at the top of the instrument is defined as the incident direction, while for an upward-going particle, the extrapolated trajectory direction at the bottom of the instrument is used.

\begin{figure}[!htb]
    \centering
    \begin{subfigure}[b]{0.45\textwidth}  
        \centering  
        \includegraphics[width=\linewidth,trim=3.0cm 0cm 6.0cm 0cm, clip]{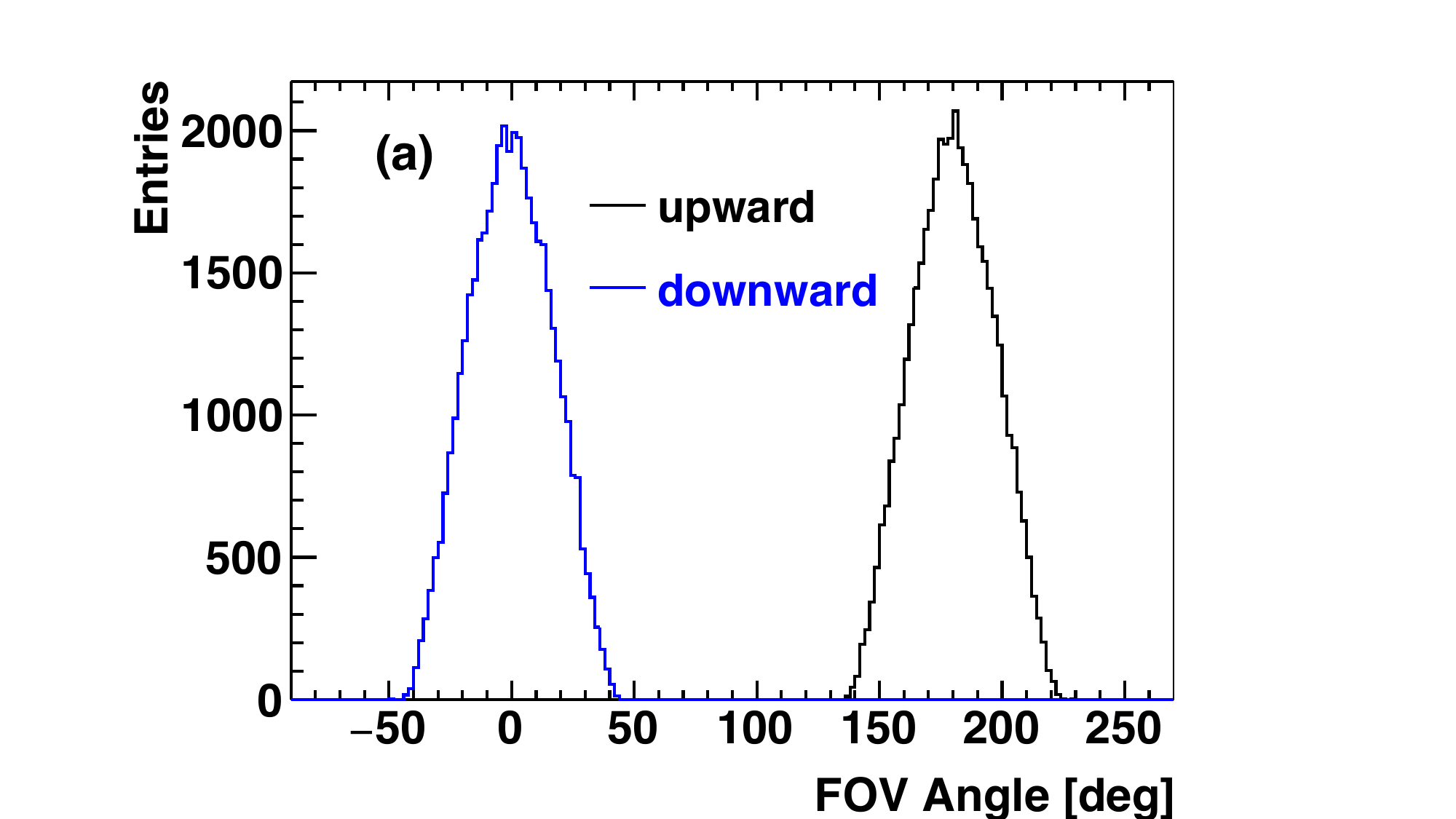}  
    \end{subfigure}
    \hfill
    \begin{subfigure}[b]{0.45\textwidth}
        \centering
        \includegraphics[width=\linewidth, trim=3.0cm 0cm 6.0cm 0cm, clip]{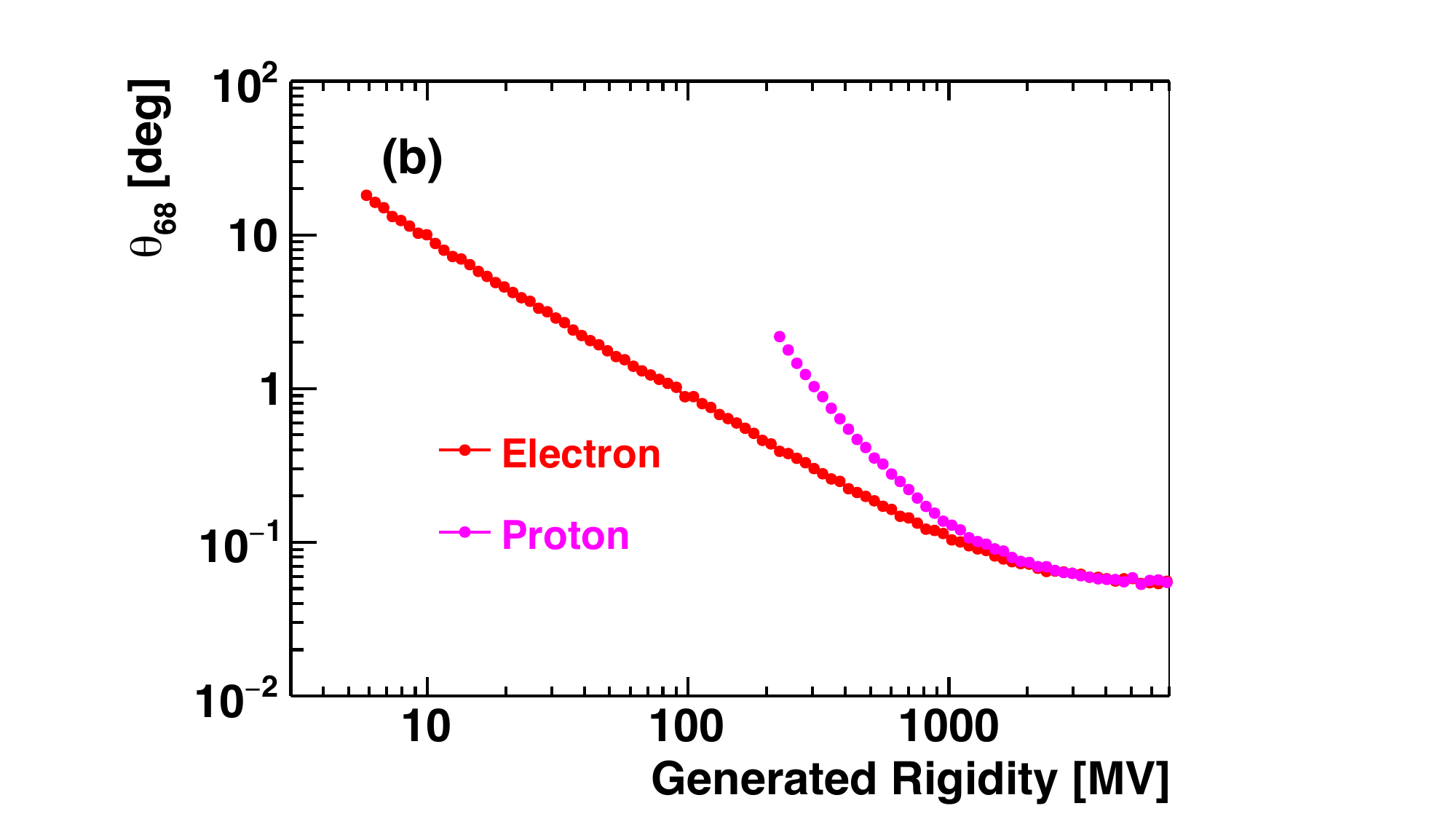}
    \end{subfigure}
    
    \caption{(a) The zenith angle distribution of downward- (blue) and upward-going (black) particles at 1 GV.  (b) The angular resolution of electrons (red) and protons (pink) as a function of generated rigidity. }
    \label{fig:angular_performance}
    
\end{figure}

Figure \ref{fig:angular_performance}(a) shows the distribution of the zenith $\theta$ of the particles in the detector acceptance. As seen, the field of view (FOV) extends to $\pm 45^\circ$ for downward-going particles, and $135^\circ - 225^\circ$ for upward-going particles.

Figure \ref{fig:angular_performance}(b) shows the angular resolution (68\% containment angle $\theta_{68}$) by comparing the reconstructed directions and the truth direction in MC simulation. As seen, the angular resolution improves with particle rigidity and approaches to $0.06^\circ$ above 1000 MV.

\subsection{Particle Identification}
With the combined information of rigidity $R$, velocity $\beta$, energy deposition $\Delta E$, and Cherenkov photon, the spectrometer is able to identify positrons, electrons, protons, helium, and other nuclei. 

\subsubsection{Separation between Positrons and Electrons}

\begin{figure}[!htb]
   \centering
    \includegraphics[width=0.75\textwidth]{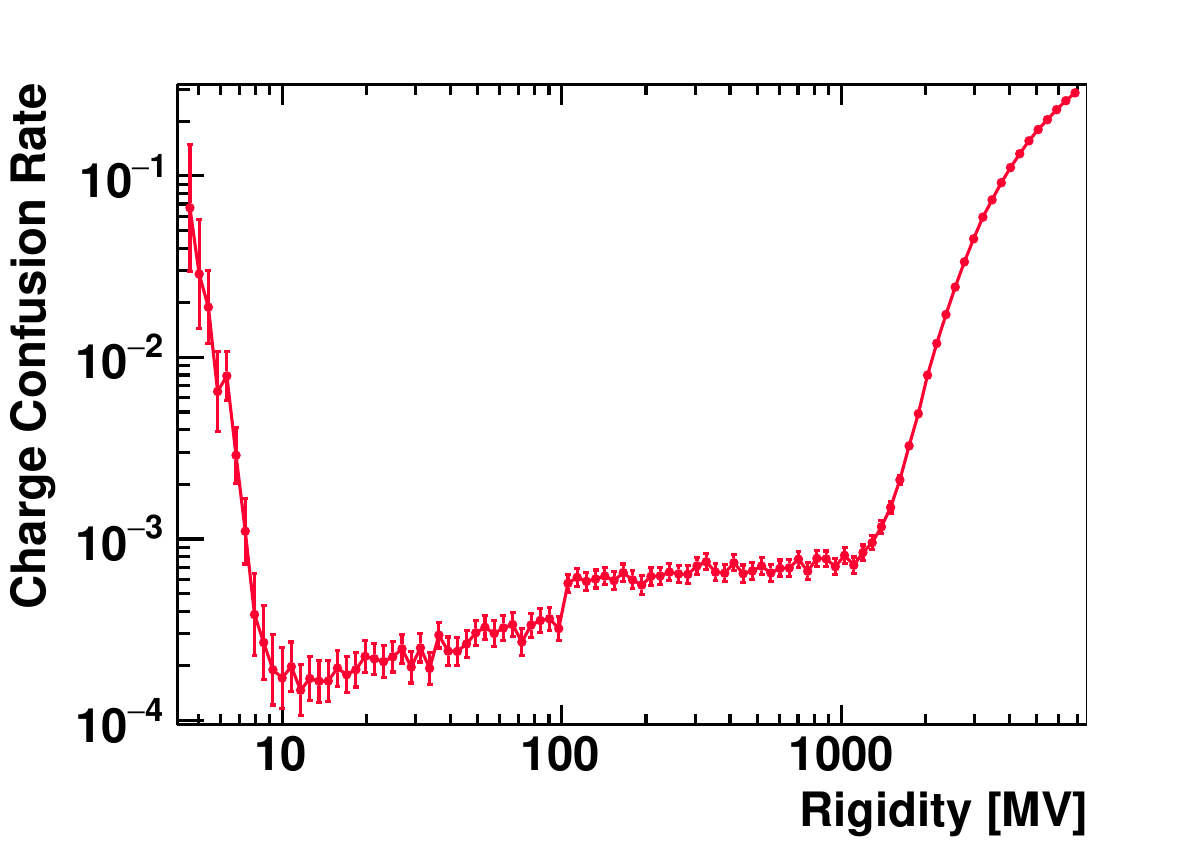}
    \caption{\footnotesize The charge sign confusion ratio of electrons, which are misidentified as positrons from the velocity direction and bending curvature. The charge confusion ratio is $<10^{-3}$ for rigidities from 8 MV to 1.3 GV. 
    }
  \label{fig:cc}
\end{figure}

Antimatter and matter particles are separated using the charge sign, which is determined from the velocity vector and rigidity. 
The charge sign could be mis-reconstructed (an electron being identified as a positron, or a proton as an antiproton) due to the limited coordinate resolution of the SciFi and STK, and the electron scattering with detector material. 
Figure \ref{fig:cc} shows the expected charge confusion probability of electrons. 
As seen, the charge sign confusion is lower than $10^{-3}$ at energies between $\sim8$ MeV and $\sim1.3$ GeV for electrons.

\subsubsection{Separation between Positrons and Protons}

Protons are the most abundant component in space radiation and are the main source of background for positron measurements. The proposed spectrometer identifies positrons using the information of $\beta$, $\Delta E/\Delta x$, and the number of Cherenkov photons in TCC. 

\begin{figure}[!htb]
   \centering
    \includegraphics[width=0.75\textwidth]{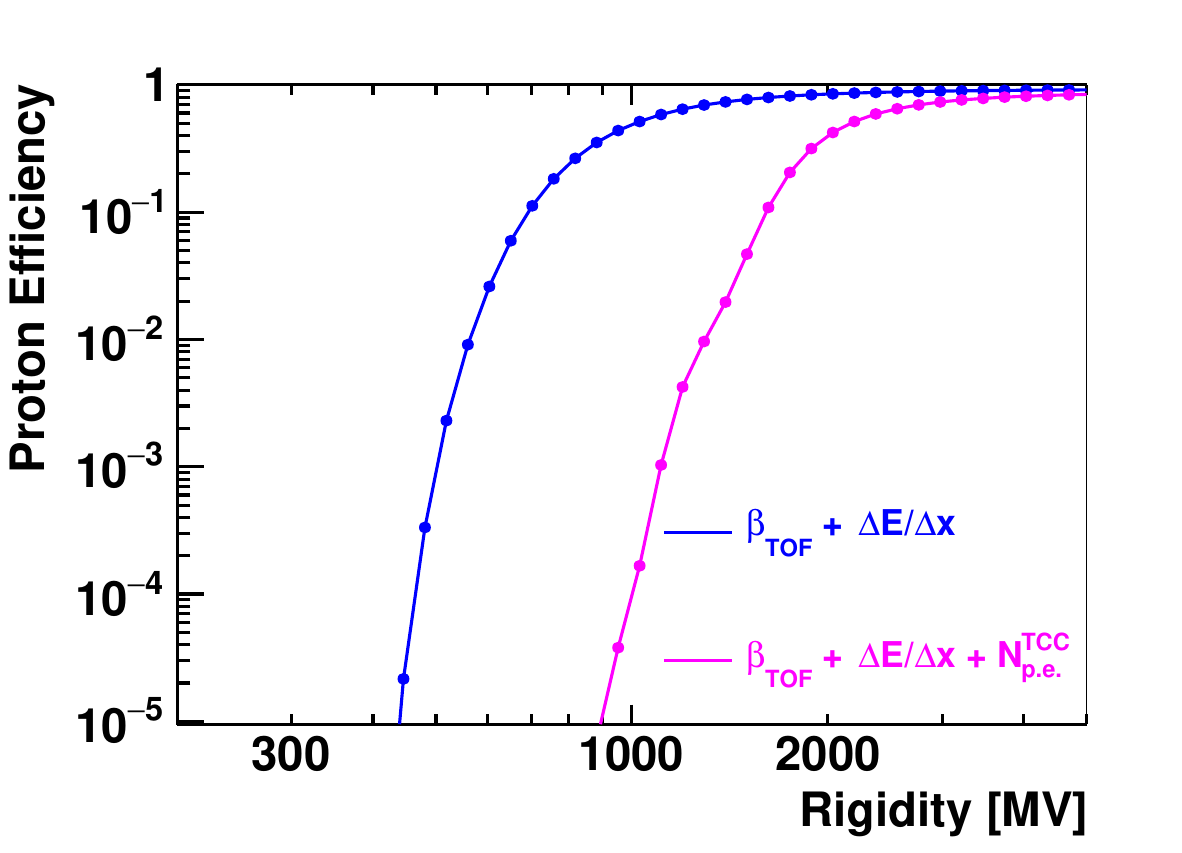}
    \caption{\footnotesize The proton rejection at $95\textrm{\%}$ positron efficiency as a function of rigidity using the $\beta_\textrm{TOF}$ + $\Delta E/\Delta x$ cut (blue), and with additional cut on $N^{\textrm{TCC}}_{\textrm{p.e.}}$ (pink).  
    }
  \label{fig:protonrej}
\end{figure}

Figure \ref{fig:protonrej} shows the expected proton rejection factor with different techniques. As seen, when combining $\beta$, $R$, and $\Delta E/\Delta x$, the proton rejection factor is better than $10^{-4}$ at 95\% positron selection efficiency for rigidities below 500 MV. With the TCC information, the proton rejection factor can be improved and maintains below $10^{-4}$ to a higher rigidity of 1 GV. 

\subsubsection{Separation between Protons and Helium}

\begin{figure}[!htb]
   \centering
    \includegraphics[width=0.75\textwidth]{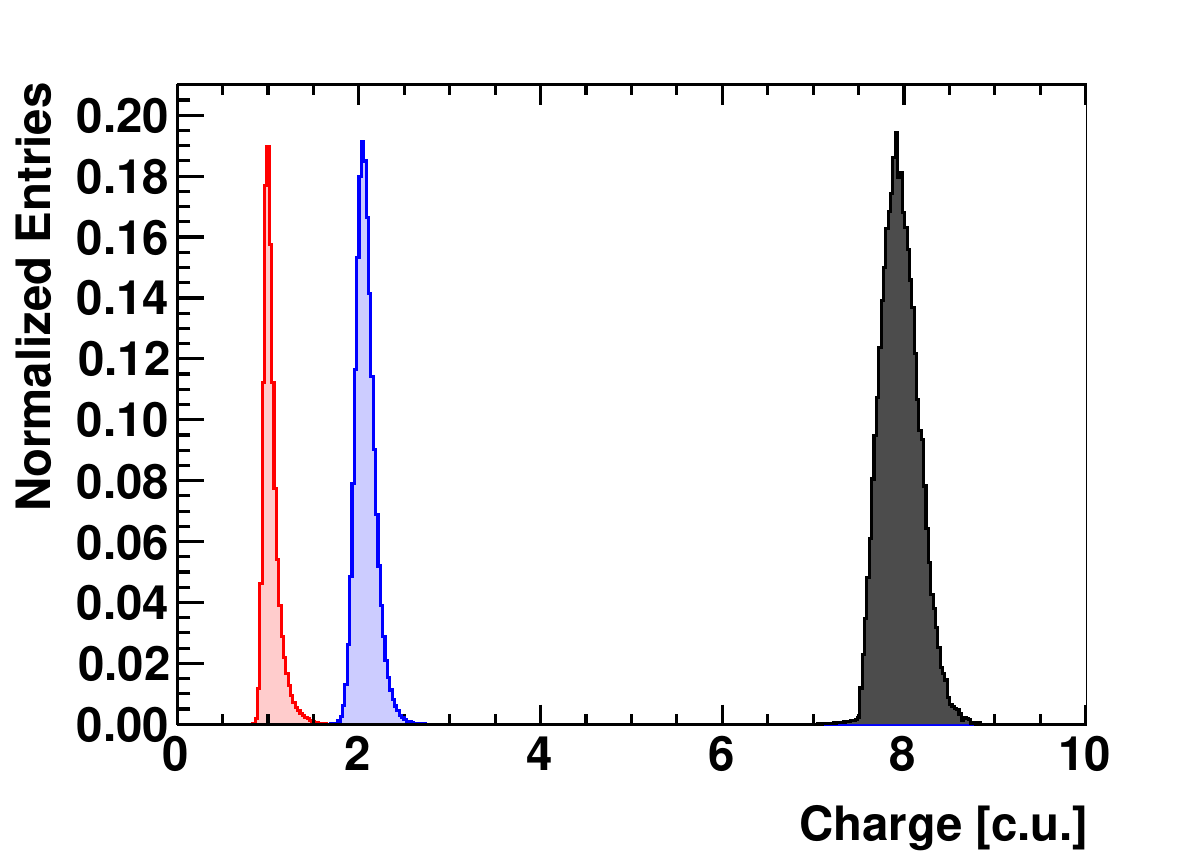}
    \caption{\footnotesize The charge resolution function for proton (red), helium (blue) and oxygen (black) at the rigidity of 3.5 GV.
    }
  \label{fig:charge_resolution}
\end{figure}

The charge of the incident particle is reconstructed from the energy depositions per unit path-length and corrected for $\beta$ or $R$ dependence. The distribution of the reconstructed charge for protons and helium at the rigidity of 3.5 GV is shown in Figure~\ref{fig:charge_resolution}. As seen, the charge resolution is 0.07 charge unit (c.u.) for protons,  0.12 c.u. for helium and 0.30 c.u. for oxygen. 

\section{Event Rate Estimation}

\begin{figure}[!htb]
   \centering
    \includegraphics[width=0.85\textwidth]{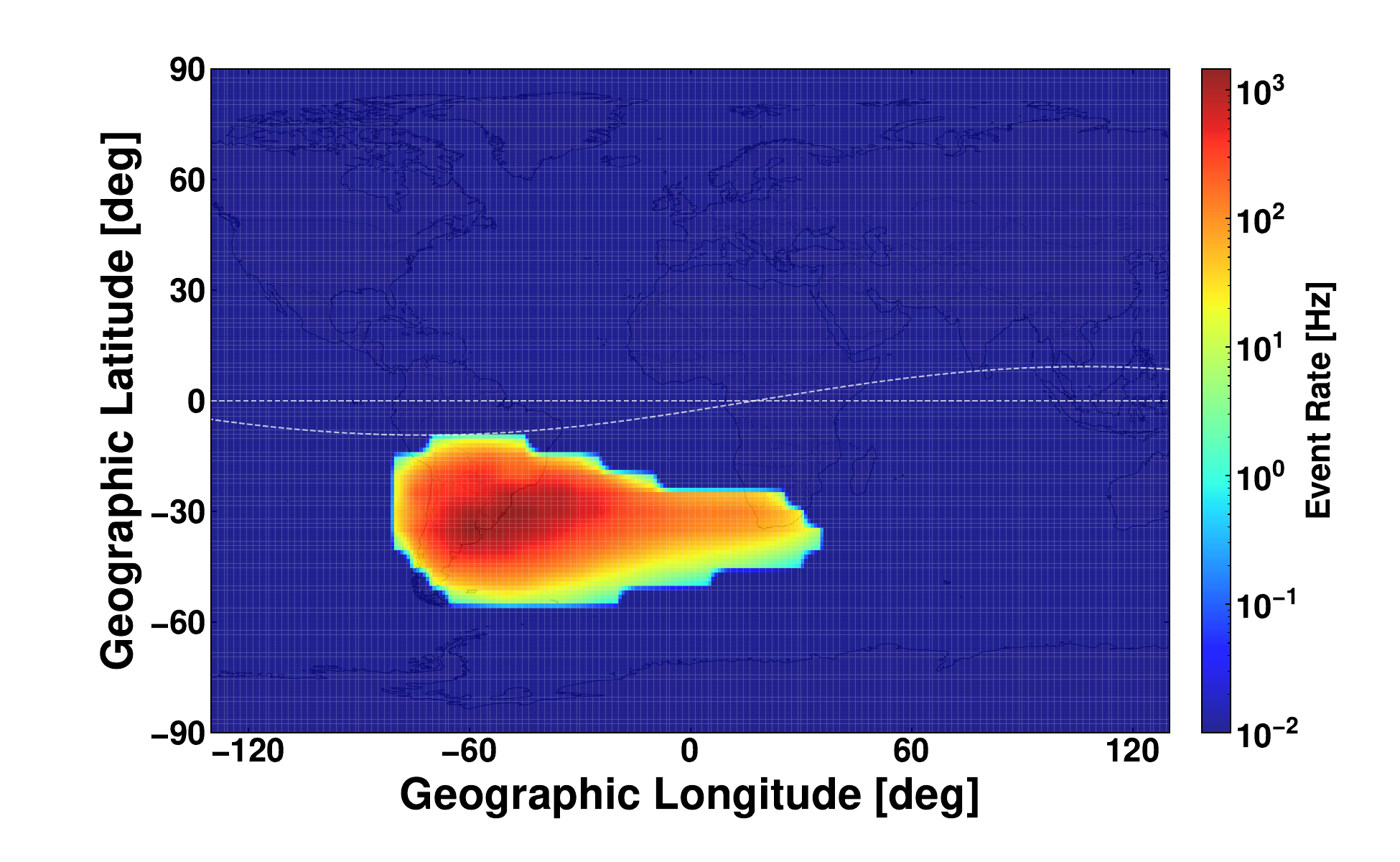}
    \caption{\footnotesize Geographical distribution of the expected proton event rate in Low Earth Orbit (LEO) at an altitude of 400 km based on AP8 model. }
  \label{fig:trigger_rate}
\end{figure}

The expected proton event rate is calculated using the proton radiation model AP8 \cite{sawyer1976ap8} and is shown in Figure~\ref{fig:trigger_rate}. As seen, the event rate is approximately $\sim1800$ Hz in the central region of the South Atlantic Anomaly (SAA).

A event rate at the kHz level is compatible with modern front-end electronics, and data acquisition systems.
The proton event rate can be further reduced by using the information from TCC to reject or prescale the protons below $\sim 1$ GV.

\newpage
\section{Conclusion}
We have presented the reference design of COMIRA, a compact magnetic spectrometer designed for precision measurements of the energy spectra and angular distributions of individual radiation component in low Earth orbit.
The detector has an overall dimension of approximately $12 \times 12 \times 12~\mathrm{cm}^3$ and an acceptance of $\sim6.6 ~\mathrm{cm}^2\mathrm{sr}$. 
By combining the measurements of particle charge, velocity, and rigidity, COMIRA can simultaneously identify and measure the electron, positron, proton, helium and other radiation components. 
The energy coverage is $\sim5$ MeV to $\sim5$ GeV for electrons,
$\sim5$ MeV to $\sim1$ GeV for positrons,
$\sim30$ MeV/n to $\sim4$ GeV/n for protons,
tens of MeV/n to GeV/n for helium and other nuclei.
The estimated event rate for a 400 km altitude low Earth orbit is approximately $\sim1800$ Hz inside the South Atlantic Anomaly.
The design can be further optimized according to the available resources of the space platform.

\section*{Acknowledgments}

This work was supported by the National Science Foundation of China (Grant No. 12505225 and 12275158).

\bibliographystyle{elsarticle-harv}

\bibliography{ref}

\end{document}